\documentclass[twocolumn,showpacs,preprintnumbers,nofootinbib]{revtex4-1}
\usepackage{graphicx,amsmath,amssymb,xcolor,hyperref}
\usepackage[utf8]{inputenc}
\usepackage{ulem}

\def\dslash{\rlap{\hspace{0.03cm}/}{\partial}}

\def\eDMEFT{e\textsc{DMeft}}

\newcommand{\SU}{\mathrm{SU}}

\begin{document}

\author{Florian Goertz}
\email{florian.goertz@mpi-hd.mpg.de}
\author{Karla Tame-Narvaez}
\email{karla.tame-narvaez@mpi-hd.mpg.de}
\author{Valentin Titus Tenorth}
\email{valentin.tenorth@mpi-hd.mpg.de}
\affiliation{Max-Planck-Institut f{\"u}r Kernphysik, Saupfercheckweg 1, 69117 Heidelberg, Germany}
\title{Di-jet/$\mathbf{e^+e^-}$\,+\,MET to Probe $Z_2-$Odd Mediators to the Dark Sector}

\begin{abstract}
We explore a scenario where Dark Matter (DM) couples to the Standard Model mainly via a scalar mediator ${\cal S}$ that is odd under a $Z_2$ symmetry, leading to interesting collider signatures. In fact, if {\it linear interactions} with the mediator are absent the most important DM production mechanisms at colliders could lead to final states with missing transverse energy (MET) in association with at least two fermions, such as di-jet or di-electron signatures. The framework we consider is model-independent, in a sense that it is only based on symmetry and formulated in the (extended) DM Effective Field Theory (\eDMEFT) approach. Moreover, it allows to address the smallness of first-generation fermion masses via suppressed $Z_2$ breaking effects.
From a di-jet + MET analysis at the LHC, we find rather loose bounds on the effective ${\cal S}$-${\cal S}$-DM-DM interactions, unless the mediator couples very strongly to SM fermions, while a future $e^+ e^-$ collider, such as CLIC, could deliver tighter constraints on the corresponding model parameters, given the mediator is leptophilic.
We finally highlight the parameter space that allows to produce the observed DM density, including constraints from direct-detection experiments.

\end{abstract}

\maketitle

%%%%%%%%%%%%%%%%%%%%%%%%%%%%%%%%%%%%%%%%
\section{Introduction and Setup}
\label{sec:intro}
%%%%%%%%%%%%%%%%%%%%%%%%%%%%%%%%%%%%%%%%

The origin of the dark matter (DM) observed in the universe is one of the biggest mysteries in modern physics. It is tackled by a multitude of experiments, which are currently running or in preparation and are probing very diverse energies. While experiments aiming for a direct detection (DD) of DM particles via nuclear recoil typically feature collision energies in the keV range, collider experiments, trying to directly produce DM particles, probe momentum transfers exceeding the TeV scale. Combining results from all such kinds of experiments in a single, consistent, yet general framework is important in order to resolve the nature of DM.

In \cite{Alanne:2017oqj}, such a framework to describe and compare searches at different energies was proposed, based on effective field theory (EFT), however allowing for detectable collider cross sections without relying on the problematic high energy tail of distributions \cite{Busoni:2013lha,Morgante:2014kra} and reproducing the correct relic density while avoiding a (too) low cutoff. 
To this end, in the \eDMEFT\ approach \cite{Alanne:2017oqj}, the field content was enlarged by a dynamical (pseudo\mbox{-)}scalar (and potentially light) {\it mediator} ${\cal S}$ to the dark sector, the latter being represented by a scalar or fermionic field $\chi$. Since both the mediator and the DM are assumed to be singlets under the SM gauge group, they can in principle interact via renormalizable couplings, however fully consistent interactions of the mediator with SM fermions (or gauge bosons) require $D=5$ operators due to gauge invariance -- which are not incorporated in typical simplified DM models \cite{Buckley:2014fba,Harris:2014hga,Abdallah:2015ter,Morgante:2018tiq}. In the \eDMEFT\ such couplings are included properly in the EFT framework, which is then consistently truncated at the $D=5$ order, leading to a well controllable number of new parameters and avoiding the need to stick to a specific UV completion. The inclusion of the most general set of (non-redundant) $D=5$ operators, allows in particular to consider richer new physics (NP) sectors, than just consisting of a single dark state and one mediator.

In this paper we focus on the phenomenology of the $D=5$ operator ${\cal S}^2\bar{\chi}_{\mathrm L} \chi_{\mathrm R}$, which can give rise to interesting di-jet phenomenology at colliders, as we will see below. If for example symmetries forbid the dimension four ${\cal S}\bar{\chi}_{\mathrm L} \chi_{\mathrm R}$ interaction, this coupling could in fact be the main portal to the Dark Sector, which could be missed in DD experiments, while mono-jet searches should be adjusted to take advantage of the peculiar di-fermion final state.

\subsection{General Setup}
We thus start from the effective Lagrangian of the SM field content, augmented with a fermion DM singlet $\chi$ and a real, CP even scalar mediator ${\cal S}$, including operators up to $D=5$, as presented in \citep{Alanne:2017oqj}, with the additional assumption that the coefficient of the operator ${\cal S}\bar{\chi}_{\mathrm L} \chi_{\mathrm R}$ is constrained to be negligibly small.\footnote{Otherwise, both the $D=4$ and $D=5$ terms would enter the following analysis -- however they could be disentangled using kinematic distributions.} 
For concreteness we will assume in the following that a symmetry forbids such $D=4$ interactions with the DM, where the most simple choice is assuming $S$ to be odd under a $Z_2$ parity, ${\cal S}$ \raisebox{2.1mm}{$\underrightarrow{\, \ \text{\footnotesize $Z_2$} \ \, }$}\,$-{\cal S}$, under which we take all the SM fields to be even, with the exception of the right-handed first fermion generation, which is also odd.

Beyond entertaining a new portal to the dark sector which is testable at (future) particle colliders, yet in agreement with null-results in DD so far, this scenario can also motivate the smallness of first-generation fermion masses, which are now forbidden at the renormalizable level.\footnote{\label{fn:S}Even though it would be tempting to address {\it all} flavor hierarchies with an (even more) extended scalar sector, linked to DM, this is beyond the scope of the present paper.}
Eventually, many of the terms in this modified \eDMEFT\ vanish compared to the original setup~\citep{Alanne:2017oqj}, including those with an odd power of mediators, unless they feature the right-handed up or down quark (or the corresponding electron). On the other hand, as mentioned, the SM-like Yukawa couplings of the latter fermions vanish and the corresponding masses will thus only be generated via small $Z_2$ breaking effects equipped with cutoff suppression.
The corresponding Lagrangian reads
\begin{eqnarray}
\label{eq:LEFT}
    \mathcal{L}_{\rm eff}^{{\cal S} \chi}& = & {\cal L}_{\rm SM^\prime} + \frac{1}{2}\partial_\mu {\cal S} \partial^\mu {\cal S} - \frac 1 2 \mu_{\cal S}^2 {\cal S}^2 + \bar \chi i \dslash\chi- m_\chi \bar \chi \chi \nonumber \\
    &-& \frac{\lambda_{\cal S}}{4} {\cal{S}}^4 - \lambda_{H{\cal S}} |H|^2 {\cal S}^2\\
    &-& \frac{\cal{S}}{\Lambda} \Big[(y_d^{\cal S})_i \bar{Q}_{\mathrm L}^i H d_{\mathrm R} +\!(y_u^{\cal S})_i \bar{Q}_{\mathrm L}^i\tilde{H} u_{\mathrm R} +\!(y_\ell^{\cal S})_i \bar{L}_{\mathrm L}^i H e_{\mathrm R} \nonumber  \\ 
    & & \quad +  \mathrm{h.c.}\Big]\nonumber\\[1mm]
    &-& \frac{y_\chi^{\cal S} {\cal S}^2 + y^H_\chi |H|^2}{\Lambda}\ \bar{\chi}_{\mathrm L} \chi_{\mathrm R} + \mathrm{ h.c.}\,, \nonumber
\end{eqnarray}
where $Q_{\mathrm{L}}$ and $L_{\mathrm{L}}$ are the left-handed $\SU(2)_{\mathrm L}$ quark and lepton doublets, resp., $d_{\mathrm R}$, $u_{\mathrm R}$, and $\ell_{\mathrm R}$ are the right-handed first-generation singlets, and $H$ is the Higgs doublet.\footnote{Note that in the case of a {\it CP-odd} scalar, $\tilde {\cal S}$, as a mediator, the Lagrangian \eqref{eq:LEFT} 
remains the same, up to the appearance of imaginary $i$ factors in the Yukawa couplings in the third line.}
The latter develops a vacuum expectation value (vev), $|\langle H \rangle| \equiv  v/\sqrt 2 \simeq 174$~GeV, triggering electroweak symmetry breaking (EWSB).  In unitary gauge, the Higgs field is expanded around the vev as  $H \simeq 1/\sqrt2 (0, v + h)^T$. Here, $h$ is the physical Higgs boson with mass $m_h \approx 125$\,GeV. Finally, ${\cal L}_{\rm SM^\prime}$ denotes the SM Lagrangian without the Yukawa couplings of the first generation, see Eq. \eqref{eq:Lm} below.

In contrast to the original setup, we assume the mediator to develop a small vev $|\langle {\cal S} \rangle| \equiv  v_{\cal S} \sim {\cal O}(1\!-\!10)$\,MeV, which finally generates masses for the first fermion generation. Since the resulting mixing with the Higgs via the $|H|^2 {\cal S}^2$ operator is suppressed, the latter  will not be considered in the following. Finally, also the "usual" dark matter coupling $\cal{S} \bar{\chi}\chi$ is generated by the spontaneous breaking of the $Z_2$-symmetry, with coefficient $\sim 2 y_\chi^{\cal S} v_{\cal S}/\Lambda$, which is however highly suppressed and only plays a role in direct detection experiments, see below.
The coefficient of the potential second $D=5$ portal to the dark sector allowed by the symmetry, $|H|^2 \bar{\chi}_{\mathrm L} \chi_{\mathrm R}$, will on the other hand taken to be small from the start, as motivated to evade direct detection constraints (remember that $v/v_{\cal S} \sim {\cal O}(10^{4})$) and limits from invisible Higgs decays (for light dark matter)\cite{Fedderke:2014wda}, playing therefore no role in the collider discussion.

%On the one hand, direct detection bounds with a CP-odd scalar mediator are much weaker due to momentum suppression of the cross section~\cite{Boehm:2014hva,Arina:2014yna}. 

Neglecting leptons for simplicity, which can be treated analogously, the resulting mass terms read
\begin{equation}
	{\cal L} \supset - \sum_{q=u,d} \bar q_L \frac{v}{\sqrt 2} \left( Y_q^H  + \frac{v_{\cal S}}{\Lambda} Y_q^S \right) q_R \equiv - \sum_{q=u,d} \bar q_L M^q q_R \,,
	\label{eq:Lm}
\end{equation}
where $q=u,d$ are three-vectors in flavor space and the Yukawa matrices
\begin{equation} 
\label{eq:Yukawas}
	Y_q^H = \begin{pmatrix}
	0 & y_{12}^q & y_{13}^q \\
	0 & y_{22}^q & y_{23}^q \\
	0 & y_{32}^q & y_{33}^q
	\end{pmatrix}\,,\quad	
	\quad Y_q^S =  \begin{pmatrix}
	({y_q^{\cal S}})_1 & 0 & 0 \\
	({y_q^{\cal S}})_2 & 0 & 0\\
	({y_q^{\cal S}})_3 & 0 & 0
	\end{pmatrix}
\end{equation}	
reflect the $Z_2$ assignments. Without breaking of the latter symmetry via $v_{\cal S}>0$, one quark family would remain massless, corresponding to a vanishing eigenvalue of $Y_q^H$. On the other hand, a small breaking of $v_{\cal S} \sim {\cal O}(10)$ MeV is enough to generate appropriate $m_u \sim m_d \sim 5$ MeV with ${\cal O}(1)$ Yukawa couplings and $\Lambda \gtrsim 1\,$TeV.

After performing a rotation to the mass basis
\begin{equation}
\begin{split}
	M^u &= U_L^u\, M_{\rm diag}^u U_R^{u\, \dagger},
	\ \, M_{\rm diag}^u\!={\rm diag}(m_u,m_c,m_t)\,,\\
	M^d &= U_L^d\, M_{\rm diag}^d U_R^{d\, \dagger},
	\ \, M_{\rm diag}^d\!={\rm diag}(m_d,m_s,m_b)\,,
\end{split}
\end{equation}
with $U_L^d = U_L^u\, V_{\rm CKM}$, we obtain the couplings of the physical quarks to the Higgs boson and the scalar mediator $\hat{Y}_q^s = U_L^{q\, \dagger}  Y_q^s U_R^q,\, s=H,S;\,q=u,d$, entering the interaction Lagrangian
\begin{equation}
	{\cal L} \supset - \sum_{q=u,d} \bar q_L \left(  \frac{\hat{Y}_q^H + v_{\cal S}/\Lambda\, \hat{Y}_q^S}{\sqrt 2}\, h + \frac{v\, \hat{Y}_q^S}{\sqrt 2 \Lambda}\, {\cal S} \right) q_R\,,
\end{equation}
where in particular the latter are crucial to test the ${\cal S}^2 \chi^2$ operator at colliders, relying on a coupling of the mediator to the SM.

\subsection{Flavor Structure}

To fully define the model, we need to fix a flavor structure, avoiding excessive flavor-changing neutral currents (FCNCs). The latter are generically generated since the fermion mass matrices $M^q$ receive contributions from different sources (see Eq.~\ref{eq:Lm}) and are in general not aligned with the individual scalar-fermion couplings $\sim Y_q^{H,S}$, such that $\hat Y_q^{H,S}$ will not be diagonal.
To this end, we first note that, in the interaction basis, the Yukawa matrices can be expressed in terms of the mass matrices as 
\begin{equation} 
\label{eq:Yukawas2}
\begin{split}
	Y_q^S  =& \frac{\sqrt 2 \Lambda}{v v_{\cal S}} \, M^q \,{\rm diag}(1,0,0) \\
	=& \frac{\sqrt 2 \Lambda}{v v_{\cal S}} \, U_L^q\, M_{\rm diag}^q U_R^{q\, \dagger} \,{\rm diag}(1,0,0)\,, \\[2mm]
	Y_q^H  =& \frac{\sqrt 2}{v} \,M^q \,{\rm diag}(0,1,1)\\	
	=& \frac{\sqrt 2}{v} \,U_L^q\, M_{\rm diag}^q U_R^{q\, \dagger} \,{\rm diag}(0,1,1)\,.
\end{split}
\end{equation}

In the mass basis, they become
\begin{equation} 
\begin{split}
	\hat Y_q^S =& \frac{\sqrt 2 \Lambda}{v v_{\cal S}} \, M_{\rm diag}^q \, U_R^{q\, \dagger} \,{\rm diag}(1,0,0)\,U_R^q\,, \\[2mm]
	\hat Y_q^H =& \frac{\sqrt 2}{v} \,M_{\rm diag}^q\, U_R^{q\, \dagger} \,{\rm diag}(0,1,1) \,U_R^q \,,
\end{split}
\end{equation}
where the unitary rotations of the left-handed fermion fields drop out since they share the same $Z_2$ charges and their couplings (with a fixed right-handed fermion) are thus aligned with the corresponding mass terms. This is not true for the right handed fermions, where the corresponding rotation matrices induce a misalignment and thus FCNCs. However, while it would not be possible to entertain $U_L^u = U_L^d = {\bf 1}$, since then $V_{\rm CKM} = {\bf 1} $, in conflict with observation, one can in fact choose the Yukawas matrices in Eq.~(\ref{eq:Yukawas2}), starting from $M_{\rm diag}^q$, such that $U_R^u = U_R^d = {\bf 1}$, avoiding FCNCs (whereas for our model the left handed rotations can be arbitrary with the only constraint $U_L^{u\,\dagger} U_L^d = V_{\rm CKM}$).\footnote{This approach is somewhat similar to the recently discussed pattern of 'singular alignment' \cite{Rodejohann:2019izm}.} Although a more systematic analysis of FCNCs in such a scenario would be interesting, we will just stick to the latter choice for the rest of this article, ending up with only diagonal couplings
\begin{equation} 
\begin{split}
	\hat Y_u^S  = & \,
	\frac{\sqrt 2 \Lambda}{v v_{\cal S}}  
	\,{\rm diag}(m_u,0,0)\,,\\
	\hat Y_d^S  = & \,
	\frac{\sqrt 2 \Lambda}{v v_{\cal S}}  
	\,{\rm diag}(m_d,0,0)\,,\\
	\hat Y_u^H  = &\, \frac{\sqrt 2}{v} 
	\,{\rm diag}(0,m_c,m_t)\,,\\
	\hat Y_d^H  = &\, \frac{\sqrt 2}{v} 
	\,{\rm diag}(0,m_s,m_b)\,.
\end{split}
\end{equation}

This means that the second and third generation couple to the Higgs boson as in the SM while the first generation couples instead only to the DM mediator, with strength determined by the free parameter $v_{\cal S}$, which we will trade for $y_u^{\cal S}/\Lambda \equiv (\hat Y_u^S)_{11}/\Lambda$ in the following. While the latter should not be too tiny, since then a very large $Z_2$-breaking vev $v_{\cal S}$ will be required to reproduce the quark masses, as discussed, ${\cal O}(1)$ values of $y_u^{\cal S} v/\Lambda$ are in perfect agreement with a modest vev and a reasonable cutoff.
 
So far we did not include the lepton sector, however a similar setup is possible for the latter, leading straightforwardly to
\begin{equation} 
\begin{split}
	\hat Y_e^S  = & \,
	\frac{\sqrt 2 \Lambda}{v v_{\cal S}}  
	\,{\rm diag}(m_e,0,0)\,,\\
	\hat Y_e^H  = &\, \frac{\sqrt 2}{v} 
	\,{\rm diag}(0,m_\mu,m_\tau)\,.
\end{split}
\end{equation}

Finally, expressing everything in terms of $y_u^{\cal S}$, we obtain the relations
\begin{equation} 
	y_e^{\cal S}\,=\, 0.1\,y_d^{\cal S}\,=\, 0.2\,y_u^{\cal S}
\end{equation}
for the couplings of the mediator to SM fermions, plugging in the values $m_u=2.5\,{\rm MeV}, m_d=5\,{\rm MeV}, m_e = 0.5\,{\rm MeV}$. As mentioned, $y_u^S/\Lambda$ can be chosen basically freely, however should not violate perturbativity of the EFT (and of the potential UV completion), which constrains $y_f^{\cal S} v/(\sqrt 2 \Lambda) < 4 \pi$\, \ [$y_f^{\cal S} < (4 \pi)^2$], for $f=u,d,e$, where we made use of the fact that the ${\cal S}-$Yukawa scales like $y_f^{\cal S} \sim g_{\rm UV}^2$.

\subsection{Relevant Parameters}
\label{sec:para}

In the following, we will derive the prospects to constrain the $Z_2$-symmetric bi-quadratic portal ${\cal S}^2\bar{\chi}_{\mathrm L} \chi_{\mathrm R}$ and the ${\cal S}$-Yukawa coupling from LHC and future ($e^+e^-$) collider data,  meeting constraints from DD and the observed relic density. A unique process where the new portal enters is fermion-pair-associated DM production, as induced by the Feynman diagrams given in Fig.~\ref{fig:FeynmannDia}, with the DM leading to a characteristic missing energy signature.
Before moving there, we will however summarize the relevant physical parameters in the model at hand. These are
\begin{itemize}
\item{the DM mass $m_{\chi}$}
\item{the mediator mass  $m_{\cal S}=\!\sqrt{\mu_{\cal S}^2\!+\!3\lambda_{\cal S} v_{\cal S}^2}$}
\item{the bi-quadratic portal coupling $y_{\chi}^{\cal S}/\Lambda$}
\item{the ${\cal S}-$Yukawa coupling $y_u^{\cal S}/\Lambda$}\,,
\end{itemize}
where we neglected potential scalar mixing from $\lambda_{H{\cal S}}$. \footnote{In the following analysis, we will consider the mediator to be much heavier than its vev, which requires an additional contribution to the Lagrangian \eqref{eq:LEFT}. While a cubic term needs a very large (non-perturbativ) coefficient, a straightforward possibility is to add another singlet ${\cal S}_2$, already envisaged in footnote \ref{fn:S}, with a ${\cal O}({\rm TeV}^2)$ quadratic term and a mass mixing ${\cal S}{\cal S}_2$ with  ${\cal O}$(1\,GeV$^2$) coefficient and/or a ${\cal S}{\cal S}_2^3$ portal with coefficient ${\cal O}(10^{-6})$. We have checked that other effects of the new scalar can be effectively decoupled.}

While this defines the main model being studied in the following sections, there are also two interesting variants obtained by either assigning positive $Z_2$ parity to all leptons or to all quarks. This will lead to a {\it leptophobic} or {\it hadrophobic} mediator, respectively, with $y_e^{\cal S} = 0$ and finite $y_d^{\cal S}=2 y_u^{\cal S}$ or vice versa.

%%%%%%%%%%%%%%%%%%%%%%%%%%%%%%%%%%%%%%%%
\section{Jets + $E^{\rm miss}_{\rm T}$ at (HL)-LHC}
\label{sec:monoj}
%%%%%%%%%%%%%%%%%%%%%%%%%%%%%%%%%%%%%%%%

\begin{figure}
 \includegraphics[width=.22\textwidth]{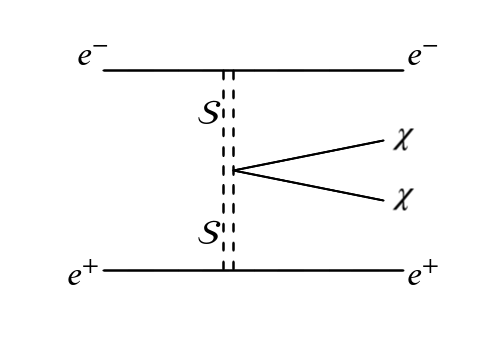}
 \includegraphics[width=.255\textwidth]{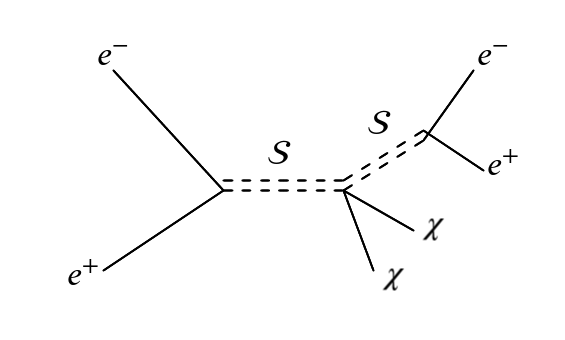}
 \caption{Feynman diagrams for dark matter + di-electron production at CLIC -- for the LHC case, the electrons are replaced by up and down quarks.}
 \label{fig:FeynmannDia}
\end{figure}

To get a first idea on near-future constraints on the new DM portal, we derive bounds from current (and projected future) LHC runs employing the CheckMate implementations of existing ATLAS analyses. A unique signature to constrain $y_{\chi}^{\cal S}$ is di-jet production in association with MET, see Fig.~\ref{fig:FeynmannDia} with the electrons replaced by up or down quarks. Here, the new portal enters at the tree-level, while the main background is $\nu \bar \nu$ production in association with jets. Although a dedicated analysis on the particular di-jet topology could improve the sensitivity, we expect the existing mono-jet search \cite{Aaboud:2017phn} using  $36.1$ fb$^{-1}$ of data and a SUSY motivated search for multiple jets plus missing energy \cite{Aaboud:2017vwy} to deliver already relevant constraints. Thus, we refrain from setting up a custom analysis but rather focus on future leptonic colliders for that purpose, where in particular the large QCD backgrounds faced at the LHC are avoided and the limits are expected to be much stronger.

Regarding the mentioned LHC analyses, the latter one naively delivers stronger constraints, but here events are used that have energies above the envisaged cutoff $\Lambda={\cal O}(1)$ TeV such that the validity is questionable \cite{Busoni:2013lha,Morgante:2014kra,Contino:2016jqw}. The scalar sum of the transverse momenta of the leading $N$ jets and $E^{\rm miss}_{\rm T}$ are required to be at least $1.6$ TeV. Therefore a reasonable value for the cut-off is at least $\Lambda\gtrsim 3$ TeV. In addition all signal-regions are inclusive ones, which means that they include events with even much higher energies, such that the resulting constraints would only be valid for borderline large couplings $y_u^{\cal S}$.

Exclusive signal regions (EM), as provided in~\cite{Aaboud:2017phn}, allow for a better estimate of the event energy. For that reason we constrain ourselves to signal regions up to EM6 of \cite{Aaboud:2017phn}, the latter containing events with $E^{\rm miss}_{\rm T}=(600-700)$\,GeV, to get robust constraints.

The signal events are simulated with MadGraph5\_aMC@NLO (v\,$2.6.5$) \cite{Alwall:2014hca}, employing a UFO \cite{Degrande:2011ua} file of our model, generated with FeynRules \cite{Alloul:2013bka, Christensen:2009jx} (to be published in the FeynRules repository). The parton-showering is done with Pythia $8.1$ \cite{Sjostrand:2006za,Sjostrand:2007gs} and the detector simulation with Delphes $3$ \cite{deFavereau:2013}, with the latter two run internally in CheckMATE $2.0.26$ \cite{Dercks:2016npn,Read:2002hq}.

\begin{figure}
    \centering
    \includegraphics[width=.42\textwidth]{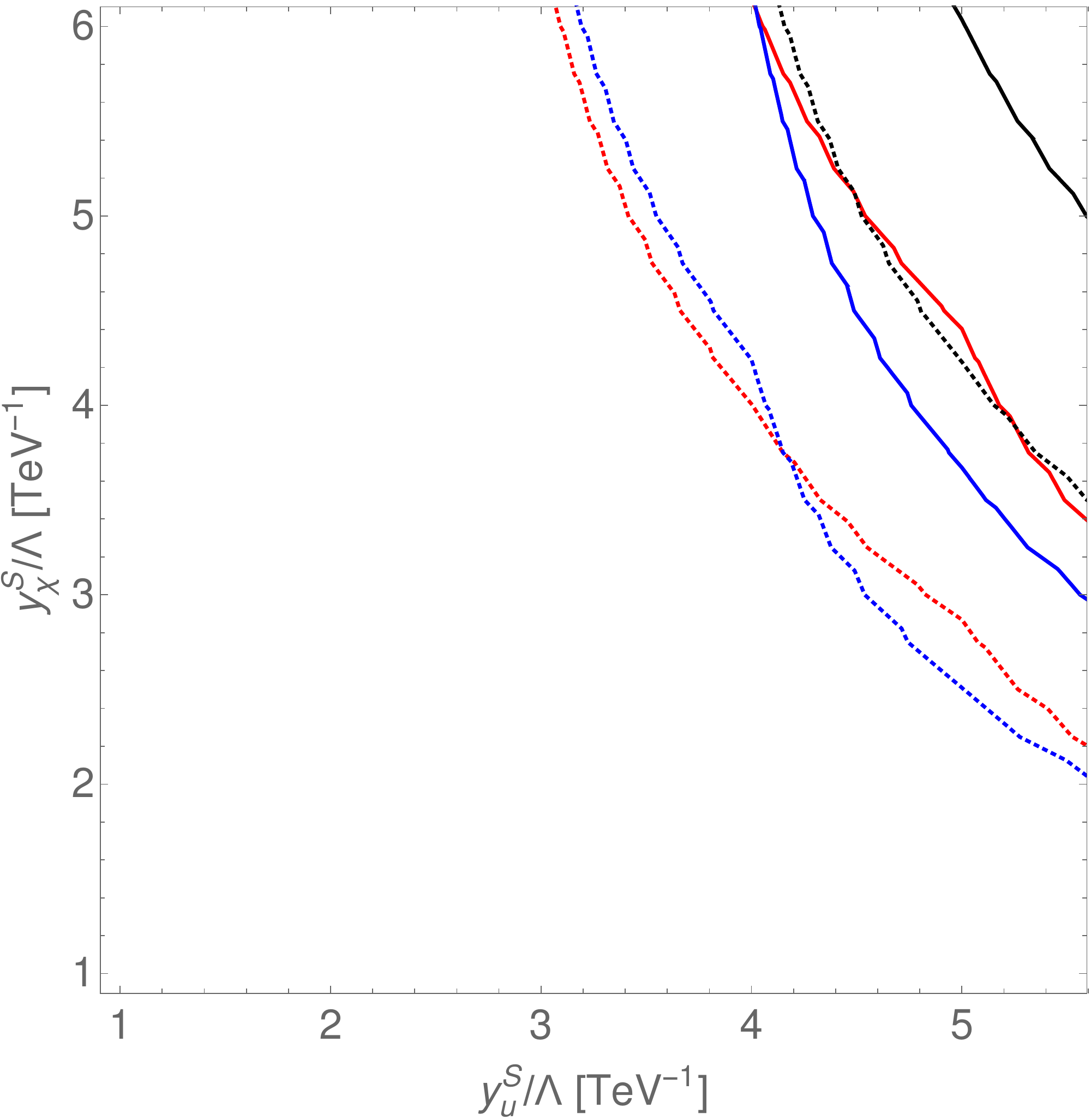}
    \caption{Exclusion reach of the current (solid) and future HL-LHC run (dotted) for $m_{\cal S}=200$ GeV and DM masses of $5$ GeV (red), $100$ GeV (blue), $300$ GeV (black).}
    \label{fig:LHC200}
\end{figure}

The actual bounds on the couplings and the prospects for the HL-LHC with a luminosity of $3$ ab$^{-1}$ are shown in Fig.~\ref{fig:LHC200} as solid and dashed lines, respectively, for $m_{\cal S}=200$\, GeV and three different DM masses, $m_{\chi}=(5, 100, 300)$\,GeV.\footnote{While with this choice the flavor model considered is fine, note that for $m_{\cal S} \gtrsim 225$\,GeV strong bounds on the ${\cal S}$-Yukawa couplings arise from the recent ATLAS search for resonant di-lepton production~\cite{Aad:2019fac}, which would exceed the projected limits of Fig.~\ref{fig:LHC200}. Clearly, this can be avoided by moving either to the leptophobic or the hadrophobic scenario.}
To obtain the projections, we used CheckMate with upscaled event numbers assuming that ATLAS measures the same distributions. Following \cite{ATL-PHYS-PUB-2018-043} we further assume that the background error can be lowered by a factor of $4$. Due to the nature of the process, radiating two DM particles from an internal mediator, interestingly the limits do not die off quickly when $m_{\chi}>m_{\cal S}/2$, allowing to test also this hierarchy of masses. As mentioned, further improvement could be reached by adjusting the analysis to the specific signature, e.g. by demanding two correlated jets in the final state. We leave the detailed study for future work.
 
We finally note that, although the final state looks similar to the one of Higgs to invisible searches in vector-boson fusion production, we found that the distribution of our signal in the main kinematic variables is very similar to the main backgrounds in that analysis and therefore no effective separation is possible there.

%%%%%%%%%%%%%%%%%%%%%%%%%%%%%%%%%%%%%%%%
\section{$e^+e^-\ +\ E^{\rm miss}_{\rm T}$ at CLIC}
\label{sec:e+e-}
%%%%%%%%%%%%%%%%%%%%%%%%%%%%%%%%%%%%%%%%

An interesting proposal for a next high-energy $e^+ e^-$ collider facility to be built is the Compact Linear Collider (CLIC) at CERN. It would be the first mature realization of a collider with these characteristics and could start running in $2035$. In the following, we will analyze the prospects to probe $y_{\chi}^{\cal S}$ at the three foreseen stages of CLIC, stage I with $\sqrt{s}=380$\,GeV, stage II with $\sqrt{s}=1.5$\,TeV and stage III with $\sqrt{s}=3$\,TeV. The corresponding luminosity goals are $1.0\,\text{ab}^{-1}$, $2.5\,\text{ab}^{-1}$, and $5\,\text{ab}^{-1}$, respectively \cite{Robson:2018enq, deBlas:2018mhx}.

To test the $Z_2$-symmetric portal we propose a search in the $e^+ e^- + E^{\rm miss}_{\rm T}$ final state at CLIC, with the signal processes depicted in Fig.~\ref{fig:FeynmannDia}. The main irreducible background is \cite{Blaising:2012vd}
\begin{equation}
    e^+e^-\ \to\ e^+e^-\bar{\nu}\nu\,,
\end{equation}
with the most important contribution coming from a $ZZ$ intermediate state, while further backgrounds turn out to be negligible.

For generating the signal and background samples at leading order, we employ again MadGraph5\_aMC@NLO for the event generation, Pythia $8.1$ for the hadronization and Delphes $3$ for a fast detector simulation. The final analysis is performed with MadAnalysis $5$ \cite{Conte:2012fm, Conte:2014zja}.

As it turns out, in the full flavor model, where $\cal{S}$ couples to electrons and quarks, the signal is very small for realistic couplings since the branching to quarks will strongly dominate (while simultaneously increasing significantly the total width).
So we first focus on the hadrophobic case, with $y_d^{\cal S}= y_u^{\cal S}=0$.\footnote{It would also be interesting to consider the di-jet final state at CLIC or to constrain the bi-quadratic portal at other colliders, however these analyses face their own challenges and will be left for future work.}

Still, we have to face a rather small signal with a sizable background, leading to weak constraints from a pure cut-and-count analysis, in particular when the uncertainty in the background cross-section normalization is taken into account. Therefore we perform a shape analysis with a binned likelihood approach, making use of the fact that our signal has a peak-like structure in the $m_{ll}$-variable -- due to an on-shell ${\cal S}$ decaying to electrons -- compared to a smoothly falling background.\footnote{In fact, the resonant diagram in the right panel of Fig.~\ref{fig:FeynmannDia} largely dominates the cross section.}

\begin{figure}
 \includegraphics[width=.45\textwidth]{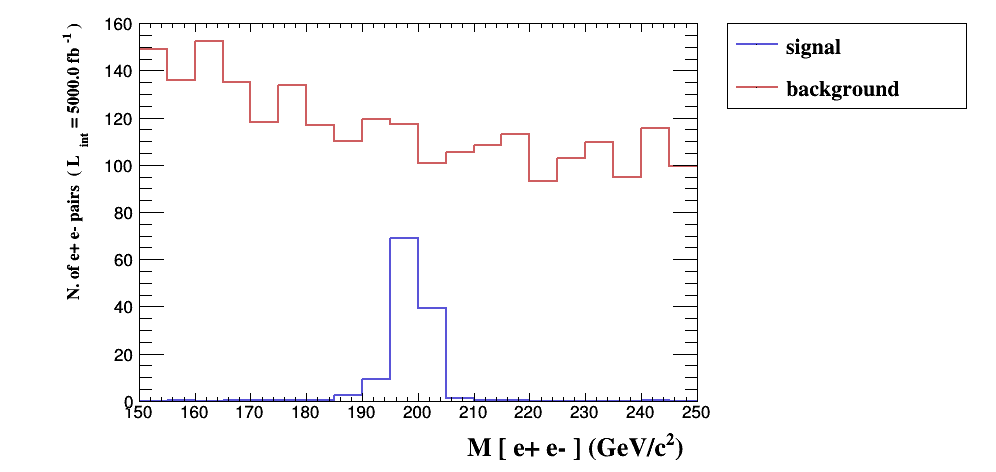}
 \caption{Comparison of signal and background shape for CLIC stage III. The signal events correspond to $y^{\cal S}_e/\Lambda=1.5$/TeV and $y^{\cal S}_\chi/\Lambda=0.25$/TeV, close to the exclusion limit.}
 \label{fig:Histo}
\end{figure}

To achieve a preliminary separation between signal and background, we apply the cuts given in Tab. \ref{tab:eecuts}, where the $m_{e^+e^-}$ cut is applied to lower the impact of $Z$ decays.
In Fig. \ref{fig:Histo} examples of the shapes of signal and background after cuts and before fitting are shown for stage III. Here, the couplings $y^{\cal S}_e/\Lambda=1.5$/TeV and $y^{\cal S}_\chi/\Lambda=0.25$/TeV are chosen to be close to the exclusion limit (see below).

\begin{table}
 \begin{tabular}{c|c|c|c|c|c}
    MET & $m_{e^+e^-}$ & $p_{T}(e)$ & $\Delta R(e^+e^-)$ & $\theta(e^+)$ & $\theta(e^-)$\\
	$[$GeV$]$ & $[$GeV$]$ & $[$GeV$]$ &  &  &  \\\hline
	$ >80$ & $>150$ & $>25$ & $<3.25$ & $>0.6$ & $<2.4$
 \end{tabular}
 \caption{Cuts for the signal region for all CLIC stages applied throughout our analysis.}
 \label{tab:eecuts}
\end{table}

\subsection{Fitting Signal and Background}

In order to use the $m_{ll}$ spectrum to discriminate signal and background, we first need sizable Monte-Carlo samples of both processes, where we generate $50.000$ and $10^6$ events, respectively. Since the signal shape depends on the width of ${\cal S}$, it is simulated for various values of the latter, depending non-trivially on the input parameters (basically $m_{\cal S}$ and $y_e^{\cal S}/\Lambda$) given at the end of Section \ref{sec:intro}.
The resulting histograms are fitted to a $4$-th order polynomial for the background and a simple Breit-Wigner distribution for the signal. Finally, the signal is characterized by the total number of events and the width of the Breit-Wigner distribution, allowing to easily test several couplings.

\subsection{The Likelihood Function}

To derive exclusion regions, we start with a binned Likelihood function \cite{Cowan:2010js} for the number of events $n_i$, similar to the one used in CheckMate \cite{Dercks:2016npn},
\begin{equation}
	{\textit L}(\mu, \theta_S, \theta_B) =\!\prod_{i} \frac{[\phi(\mu, \theta_S, \theta_B)]^{n_i}\!}{n_i!}\, e^{\!-\phi(\mu, \theta_S, \theta_B)} e^{\!-\theta_S^2/2-\theta_B^S/2}\,,
\end{equation}
with
\begin{equation}
	\phi(\mu, \theta_S, \theta_B)\ =\ \mu S e^{\sigma_S \theta_S}\ +\ B e^{\sigma_B \theta_B}
\end{equation}
and
\begin{equation}
	\sigma_S=\frac{\Delta S}{S}\ , \quad \sigma_B=\frac{\Delta B}{B}\,.
\end{equation}

Here, $S$ and $B$ are the predicted numbers of signal and background events, respectively, while $\theta_{S,B}$ are nuisance parameters incorporating the corresponding uncertainties $\Delta S$ and $\Delta B$. Finally, the variation of the signal strength with the input parameters, given in Sec.~\ref{sec:para}, is parameterized by the signal-strength modifier $\mu$, which is normalized for fixed $y_e^{\cal S}/\Lambda$ and fixed masses such that $\mu=(y_{\chi}^{\cal S}/\Lambda)^2$. 

To test the compatibility of different values for the latter with data, we use the profile likelihood ratio \cite{Cowan:2010js}
\begin{equation}
	\tilde{\lambda}({\mu}) = \left\{ \! \! \begin{array}{ll}
       \frac{{\textit L}(\mu, \hat{\hat{\theta}}_S(\mu), \hat{\hat{\theta}}_B(\mu))}{{\textit L}(\hat{\mu}, \hat{\theta}_S, \hat{\theta}_B)}\ & \hat{\mu} \ge 0 , \\*[0.3 cm]
        \frac{{\textit L}(\mu, \hat{\hat{\theta}}_S(\mu), \hat{\hat{\theta}}_B(\mu))}{{\textit L}(0, \hat{\hat{\theta}}_S(0), \hat{\hat{\theta}}_B(0))}\ & \hat{\mu} < 0 \;\,,
         \end{array} \right.
\end{equation}
were $\hat{\hat{\theta}}_S(\mu), \hat{\hat{\theta}}_B(\mu)$ maximize ${\textit L}$ for the given value of $\mu$, while $\hat{\mu}, \hat{\theta}_S, \hat{\theta}_B$ correspond to the unconditional (global) maximum appearing in the denominator and are called unconditional Maximum Likelihood (ML) estimators. Here, the lower case accounts for the fact that we can only have a positive signal contribution.

Finally, for the numerical analysis it is convenient to use the test statistics \cite{Cowan:2010js}
\begin{equation}
	\tilde{q}_{\mu} = \left\{ \! \! \begin{array}{ll}
               - 2 \ln \tilde{\lambda}(\mu)  & \hat{\mu} \le \mu  \\*[0.2 cm]
               0 & \hat{\mu} > \mu
              \end{array} \right.
\end{equation}
to set upper limits (with higher values corresponding to less compatibility), for which we use the python package \texttt{iminuit} \cite{iminuit}.

\subsection{P-Values}

In the following we assume that the true underlying theory features $\mu=0$, i.e. we expect to see background only, and want to derive corresponding projected experimental exclusion regions on $\mu$.

In general, to quantify the agreement between a (potentially) observed measurement and a signal hypothesis $\mu\!>\!0$, leading to a certain $\tilde{q}_{\mu,obs}$, the $p-$value
\begin{equation}
	p_\mu = \int_{\tilde{q}_{\mu,obs}}^{\infty} f(\tilde{q}_\mu|\mu)d\tilde{q}_\mu
\end{equation}
is calculated, where $f(\tilde{q}_\mu|\mu^\prime)$ is the probability density function (pdf) of $\tilde{q}_\mu$ under the assumption that the data is distributed according to a true $\mu=\mu^\prime$, while the subscript in the first argument denotes the hypothesis being tested.\footnote{In fact, this quantifies the probability that, given the true signal strength is $\mu$, we will observe a value of $\tilde{q}_{\mu}$ as large as $\tilde{q}_{\mu,obs}$ (or larger).}
As we want to derive the {\it expected} upper limits from future experiments, assuming no signal to be present, we will use the median value of the corresponding distribution, $f(\tilde{q}_{\mu}|0)$, for $\tilde{q}_{\mu,obs}$. Finally, working at the $95\%$ confidence level, we will solve for the value of $\mu$ that leads to $p_\mu=0.05$.

To obtain the distributions $f(\tilde{q}_\mu|\mu^\prime)$ without a large number of Monte Carlo simulations,
we use the asymptotic formulas given in Ref.~\cite{Cowan:2010js}. Those are valid for a sufficiently high number of events in each bin, which is fulfilled in our case.\footnote{We have checked the (rough) agreement of the asymptotic formula with generated distributions for several values of $\mu$.} 
While in the case $\mu^{\prime}=\mu$, $f(\tilde{q}_\mu|\mu)$ is given by a simple half-chi-square distribution, for obtaining the median of $\tilde{q}_{\mu}$ according to $f(\tilde{q}_{\mu}|0)$ the so-called Asimov data set is used \cite{Cowan:2010js}, where all estimators obtain their true values. This data set can be approximated via large MC simulations. Here we assume that our initial sets are large enough and use the fitted distributions as Asimov data. 
With this, the corresponding Likelihood-function and test statistics can be evaluated, which are denoted by ${\cal L}_A$ and $q_{\mu,A}$. The variance, from which $f(\tilde{q}_{\mu}|0)$ can be obtained, is then simply given by $\sigma^2_A = \frac{\mu^2}{q_{\mu,A}}$, assuming background-only \cite{Cowan:2010js}.
In practice we can however just use the Asimov value $q_{\mu,A}$ for the median of $[\tilde{q}_{\mu}|0]$, according to \cite{Cowan:2010js}, and therefore the expected $p-$value for a signal hypothesis becomes
\begin{equation}
	p_{\mu} = 1 - \Phi\Big(\sqrt{q_{\mu,A}}\Big)\,,
\end{equation}
with $\Phi$ the cumulative Gaussian distribution. In the end, $p_{\mu}$ is evaluated for varying $\mu$ to find $p_{\mu}=0.05$.

\subsection{Resulting Limits}

\begin{figure}
	\includegraphics[scale=.51]{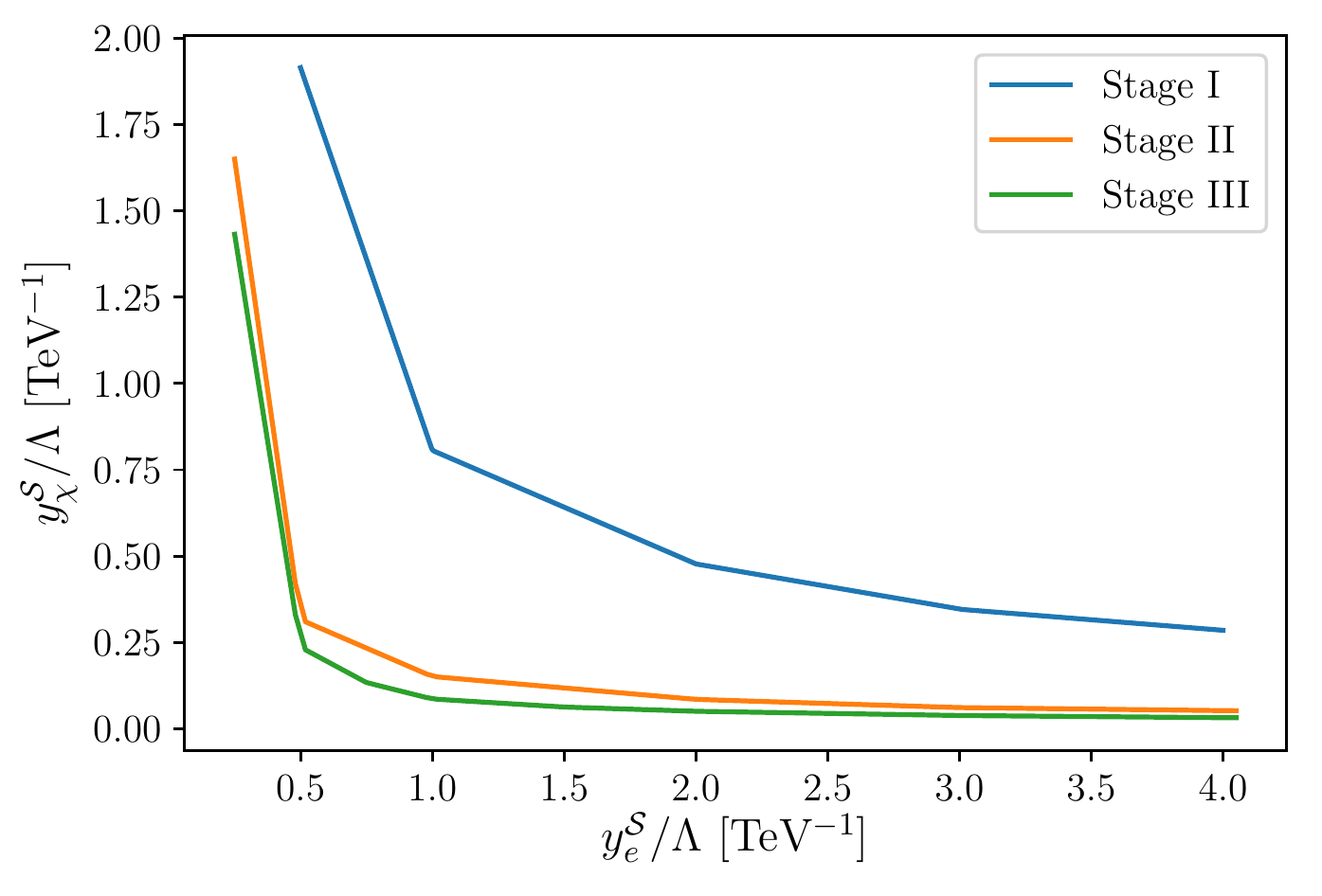}
	\caption{Comparison of the expected limits on the couplings obtained at the three stages of CLIC, assuming $m_{\cal S}=200$\,GeV and $m_\chi=5$\,GeV.}
	\label{fig:allstages}
\end{figure}

\begin{figure}
	\includegraphics[scale=.51]{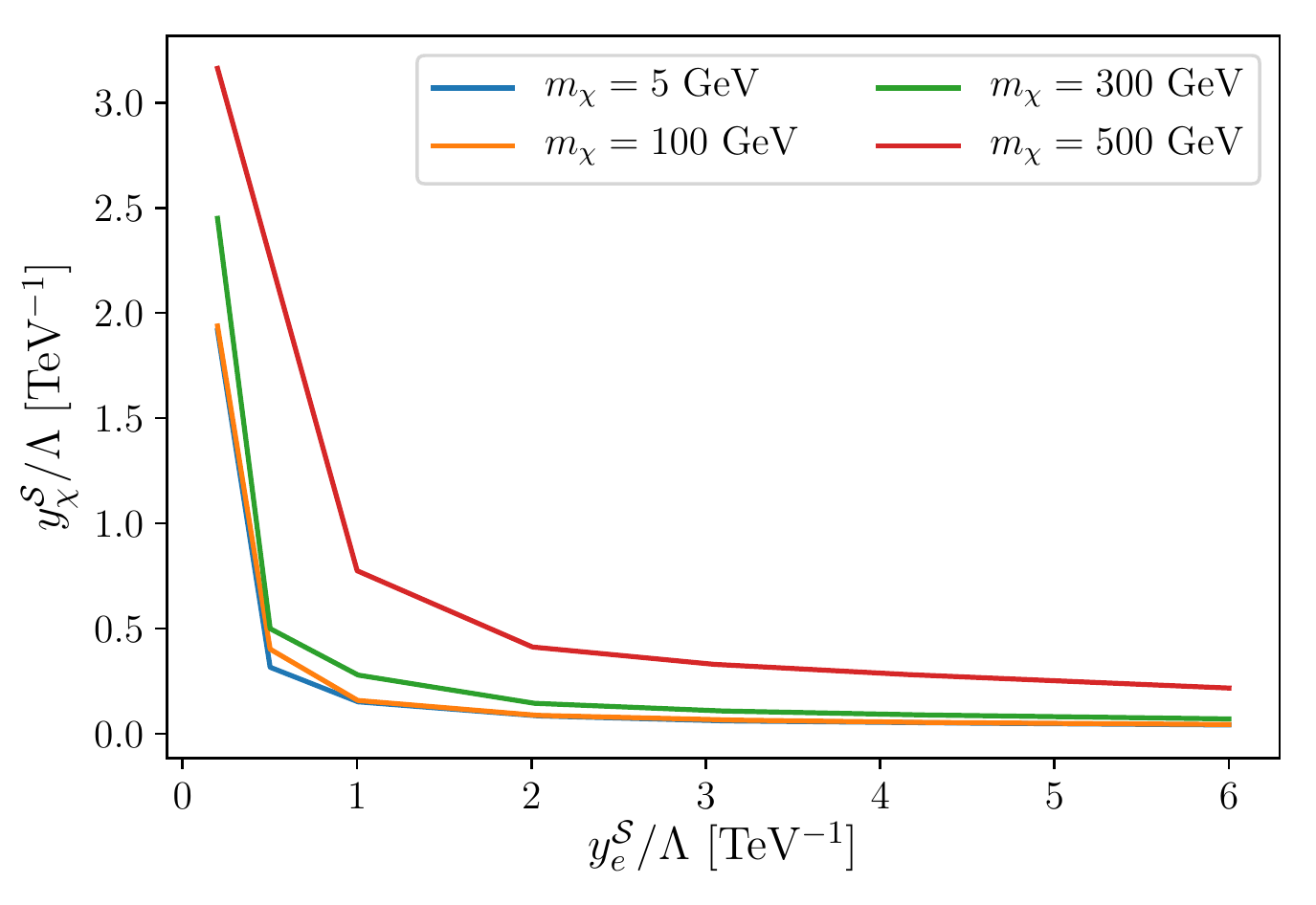}
	\caption{Expected limits on the couplings obtained at the second stage of CLIC, with $\sqrt{s}= 1.5$\,TeV, for $m_{\cal S}=200$\,GeV and several dark matter masses.}
	\label{fig:ms200}
\end{figure}

To establish the constraints on the model parameters, we have to translate the limits on $\mu$ into limits for the former.
As mentioned before, for fixed $y_{\cal S}^e$  and thereby fixed width and shape of the $m_{ee}$ distribution, we have $\mu=(y_{\chi}^{\cal S}/\Lambda)^2$. For all limits we take a $5\%$ uncertainty on the background normalization into account, i.e., $\sigma_B=0.05$ (while $\sigma_S$ is negligible).

In Fig.~\ref{fig:allstages} we compare the reach of the three CLIC stages on the couplings, assuming $m_{\cal S}=200$\,GeV and $m_\chi=5$\,GeV. We observe that already at the first stage we would be sensitive to ${\cal O}(1/{\rm TeV})$ couplings, while at the later stages the reach extends well beyond a TeV. In Fig. \ref{fig:ms200} the expected limits obtained for the same $m_{\cal S}$, but varying dark matter masses, are shown for CLIC stage II, which demonstrates that the sensitivity does not vanish for $m_\chi/2 > m_{\cal S}$.

We further note that direct searches for the mediator, e.g. in the $e^+e^-$ final state, could break the degeneracy between the two couplings. It might well happen that the mediator would first be found via such a search, however then the present analysis would be crucial to investigate the structure of the dark sector.

\begin{figure}[t]
 \includegraphics[width=.45\textwidth]{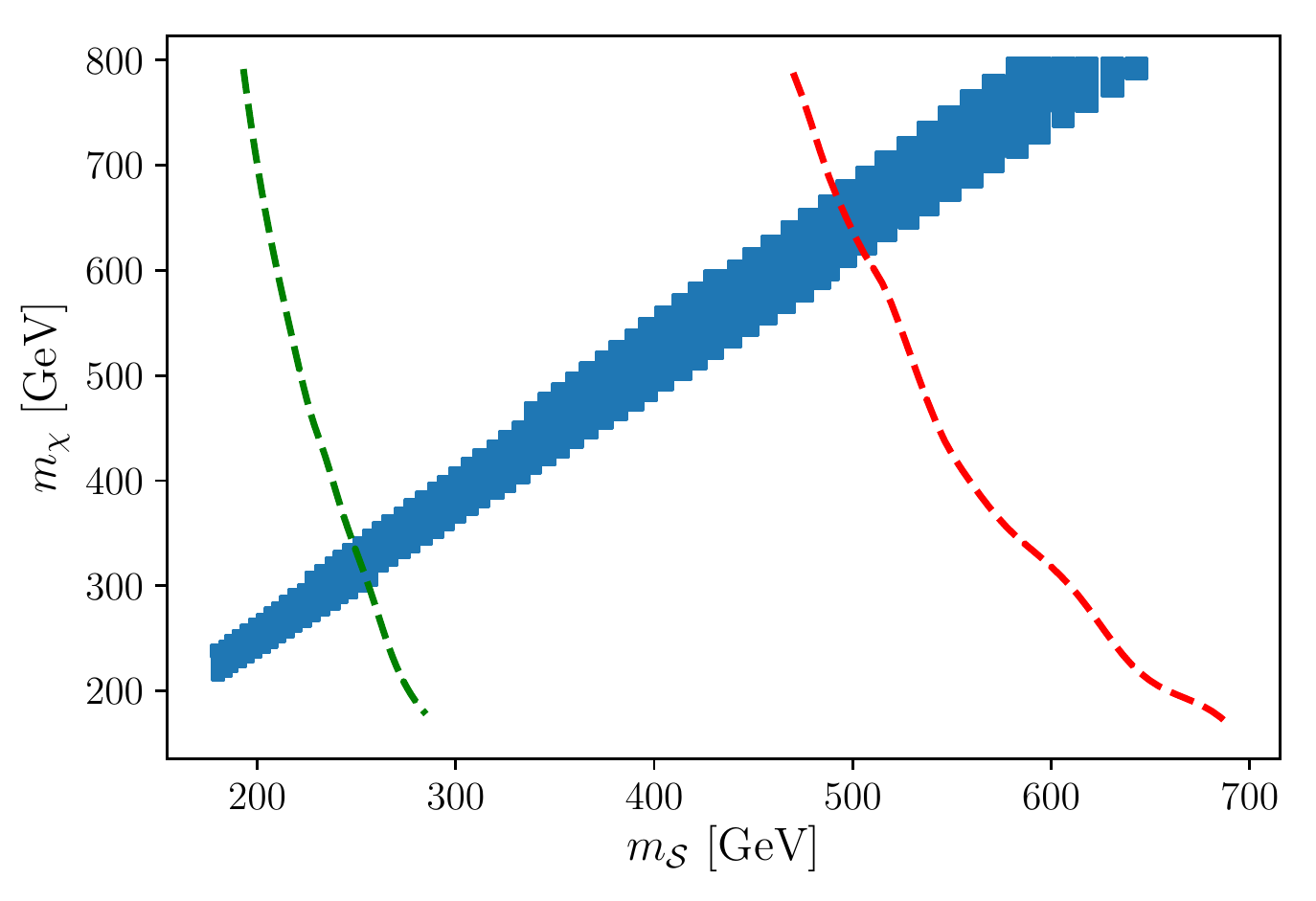}
 \caption{Band of relic density $0.11\!<h^2\Omega_{\rm DM}\!< \!0.13$ (dark~blue) for $y^{\cal S}_\chi=2.25$, independent of $y^{\cal S}_{u,d,e}$. Exclusions from XENON1t (left of green line) and the LZ projection (left of red curve) are superimposed (which however are not present for the hadrophobic model). The leftover space can be tested with DARWIN.}
 \label{fig:RDmSmchi}
\end{figure}

\begin{figure}[t]
 \includegraphics[width=.45\textwidth]{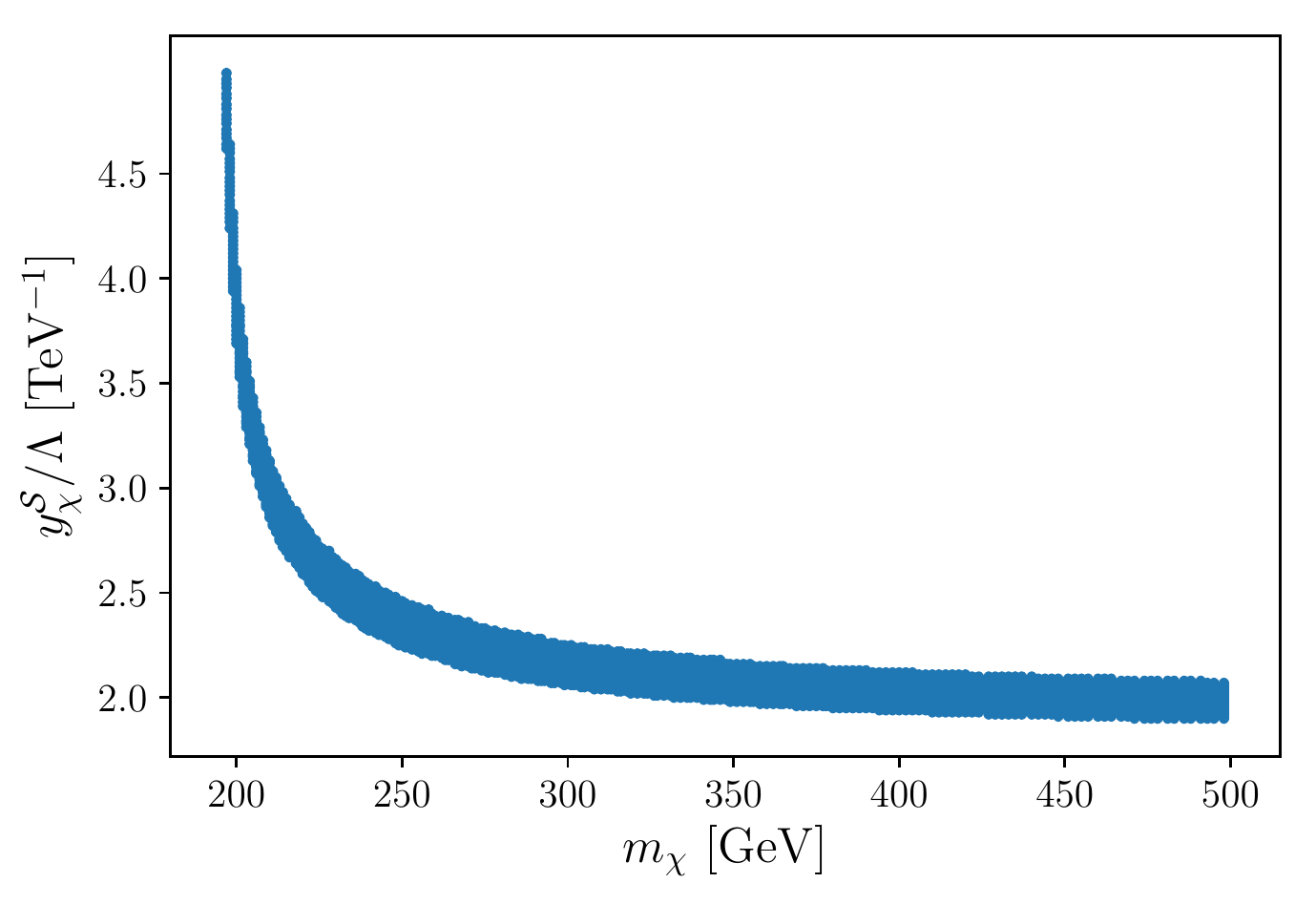}
 \caption{Band of relic density $0.11\!<h^2\Omega_{\rm DM}\!< \!0.13$ for \mbox{$m_{\cal S}=200$\,GeV}.}
 \label{fig:RDmchiyS}
\end{figure}

\vspace{.7 cm}

%%%%%%%%%%%%%%%%%%%%%%%%%%%%%%%%%%%%%%%%
\section{Dark Matter Phenomenology}
\label{sec:RD}
%%%%%%%%%%%%%%%%%%%%%%%%%%%%%%%%%%%%%%%%

For $m_\chi \gtrsim m_{\cal S}$, the DM relic density is set via the process $\bar{\chi}\chi \to {\cal S}{\cal S}$, while for smaller dark matter masses it is always far above the measured value since no decay channel is kinematically allowed (the $s$-channel decay induced by $v_{\cal S}>0$ is found to be negligible, even in the resonance region). The viable parameter region, featuring $0.11\!<h^2 \Omega_{\rm DM}\!<\!0.13$, is shown as a blue band in Fig.~\ref{fig:RDmSmchi} in the $m_{\cal S}-m_\chi$ plane, where we set $y^{\cal S}_\chi=2.25$. Light mediators $m_{\cal S} < 200$\,GeV, below the green line, are already excluded by XENON1t \cite{Aprile:2017iyp} and heavier once will be tested in future experiments like LZ \cite{Szydagis:2016few} (red line) and DARWIN \cite{Aalbers:2016jon} (remaining region). The dominant contribution to direct detection rates arises from tree-level $s$-channel exchange of $\cal{S}$ with the up and down quarks and therefore vanishes in the hadrophobic case. Since $v_{\cal S} \propto 1/y^{\cal S}_f$, the cross section is independent of the Yukawa couplings. All numerical results have been obtained with micrOmegas $5.0.8$ \cite{Belanger:2018mqt}.

Finally, the required $y_\chi^{\cal S}$ in dependence on $m_\chi$ is shown in Fig.~\ref{fig:RDmchiyS} for $m_{\cal S} = 200$ GeV. Note that also the relic density is independent of the values of $y_u^{\cal S}$ (or $y^{\cal S}_e$), which do not enter the dominant annihilation amplitude. We find that, unless the electron ${\cal S}-$Yukawa coupling is very small, most of the viable parameter space will be tested at CLIC.

%%%%%%%%%%%%%%%%%%%%%%%%%%%%%%%%%%%%%%%%
\section{Acknowledgments}
\label{sec:Ack}
%%%%%%%%%%%%%%%%%%%%%%%%%%%%%%%%%%%%%%%%

We thank Tommi Alanne, Giorgio Arcadi, Thomas Hugle, Felix Kahlhoefer, Ulises Saldaña-Salazar, and Stefan Vogl for helpful discussions.
VT acknowledges support by the IMPRS-PTFS and
KTN by a CONACYT-CONCYTEP grant.

\bibliography{Di-Jet}

\end{document}